%% file: main.tex
\documentclass[conference]{IEEEtran}
\IEEEoverridecommandlockouts
\usepackage{cite}
\usepackage{amsmath,amssymb,amsfonts}
\usepackage{algorithm}
\usepackage{algorithmic}

\usepackage{graphicx}
\usepackage{textcomp}
\usepackage{xcolor}
\usepackage{subcaption}
\usepackage{multirow}
\pdfoutput=1 
\usepackage{lipsum}
\usepackage{mathtools}
\usepackage{cuted}
\usepackage{stfloats}
\usepackage{amsthm}

\usepackage{enumitem}

\allowdisplaybreaks

\begin{document}

\title{Multi-IRS-assisted Multi-Cell Uplink MIMO Communications  under Imperfect CSI: A Deep Reinforcement Learning Approach
}
\author{\IEEEauthorblockN{Junghoon Kim\IEEEauthorrefmark{1}, Seyyedali Hosseinalipour\IEEEauthorrefmark{1}, Taejoon Kim\IEEEauthorrefmark{2}, David J. Love\IEEEauthorrefmark{1} and Christopher G. Brinton\IEEEauthorrefmark{1}}
\IEEEauthorblockA{\IEEEauthorrefmark{1}Electrical and Computer Engineering, Purdue University, West Lafayette, IN, USA}
\IEEEauthorblockA{\IEEEauthorrefmark{2}Electrical Engineering and Computer Science, University of Kansas, Lawrence, KS, USA}
\IEEEauthorblockA{\IEEEauthorrefmark{1}\{kim3220, hosseina, djlove, cgb\}@purdue.edu, \IEEEauthorrefmark{2}taejoonkim@ku.edu}}
\maketitle

\begin{abstract}

\input{abstract}

\end{abstract}

\input{intro}

\input{formulation}

\input{optimization}


\input{MADRL}

\input{sim}

\input{conc}

\section*{Acknowledgment}
D.J. Love was supported in part by the National Science Foundation (NSF) under grants CNS1642982 and CCF1816013. C. G. Brinton was supported in part by the NSF under grants AST2037864. T. Kim was supported by the NSF under grants CNS1955561.





\bibliographystyle{IEEEtran}
\bibliography{ref}

\end{document}

%% file: abstract.tex
Applications of intelligent reflecting surfaces (IRSs) in wireless networks have attracted significant attention recently. Most of the relevant literature is focused on the single cell setting where a single IRS is deployed and  perfect channel state information (CSI) is assumed. 
In this work, we develop
a novel methodology for \textit{multi-IRS-assisted  multi-cell networks} in the uplink. 
We consider the scenario in which  (i) channels are dynamic and (ii) only partial CSI is available at each base station (BS); specifically, scalar effective channel powers from only a subset of user equipments (UE).
We formulate the sum-rate maximization problem aiming to jointly optimize the IRS reflect beamformers, BS combiners, and UE transmit powers. 
In casting this as a sequential decision making problem, we propose a multi-agent deep reinforcement learning algorithm to solve it, where each BS acts as an independent agent in charge of tuning the local UE transmit powers, the local IRS reflect beamformer, and its combiners. We  introduce an efficient information-sharing scheme that requires limited information exchange among neighboring BSs to cope with the non-stationarity caused by the coupling of actions taken by multiple BSs. Our numerical results show that our method obtains substantial improvement in average data rate compared to baseline approaches, e.g., fixed UE transmit power  
and maximum ratio combining.

%% file: intro.tex
\section{Introduction}
\label{sec:intro}
 Intelligent reflecting surfaces (IRSs) are one of the innovative technologies for 6G and beyond~\cite{zhang2020prospective,hosseinalipour2020federated}. 
 An IRS is an array of passive reflecting elements with a control unit. It manipulates the propagation of an incident signal by providing an abrupt phase shift,
 which can control the communication channel.  IRSs are utilized to provide enhanced communication efficiency without building extra infrastructure~\cite{8811733,9013288,pan2020multicell,8990007,zhi2020uplink,9162705,zhang2020intelligent}. In this paper, we study a scenario where multiple IRSs are deployed in a multi-cell cellular setting to provide enhanced data rates to the users.

 \subsection{Related Work}
 Exploiting IRSs in cellular networks initiated with applications of this technology in the downlink (DL). Studying IRS use cases in the uplink (UL) is thus comparably more recent. 
\subsubsection{Utilizing IRSs in the DL} Most of the relevant literature has considered a \textit{single cell} system with a \textit{single IRS}~\cite{8811733,9013288}.
Specific investigations have included quality of service (QoS)-constrained transmit power minimization~\cite{8811733} and weighted sum-rate maximization~\cite{9013288}
to obtain the base station (BS) beamformer and IRS reflect beamformer/precoder in the DL.  
Unlike the prior approaches, the work in~\cite{pan2020multicell} considers 
a \textit{multi-cell} scenario with a single IRS, where the BS precoders and IRS reflect  beamformer are designed to maximize sum-rate. 

\subsubsection{Utilizing IRSs in the UL} 
Most of the works in UL design are also focused on single cell systems with a single IRS~\cite{8990007,zhi2020uplink,9162705}.
Several of these works have studied IRS reflect beamformer design and uplink user equipment (UE) power control problems, where the impact of quantized IRS phase values \cite{8990007} and compressed sensing-based user detection \cite{9162705} on the uplink throughput have also been investigated.
The concept of IRS resembles analog beamforming in millimeter-wave (mmWave)-based systems~\cite{zhi2020uplink}.
Recently, systems with two IRSs have been considered focusing on SINR fairness~\cite{zhang2020intelligent}.

Despite the potential benefit of improving multi-cell-wide performance, multi-IRS deployment in multi-cell UL scenarios has not been thoroughly modeled and studied due to the added optimization complexity involved in controlling multiple IRSs.

\subsection{Overview of Methodology and Contributions}


In this work, we develop an architecture for multi-IRS-assisted multi-cell UL networks. 
Our methodology explicitly considers
multi-order reflections among IRSs, which is rarely done in existing literature. 
We address the scenario where (i) channels are time-varying, and (ii) only partial/imperfect CSI is available, in which each BS only has  knowledge of scalar effective channel powers from a subset of UEs.
This is more practical and realistic as compared to the prior approaches \cite{8990007,zhi2020uplink,9162705,zhang2020intelligent} that assume  perfect knowledge of all channel matrices.
We formulate the sum-rate maximization problem aiming to jointly optimize UE transmit powers, IRS reflect beamformers, and BS combiners across cells.
%

Given the interdependencies between the design variables across different cells, we cast the problem as one of
sequential decision making and tailor a multi-agent deep reinforcement learning (DRL)
algorithm to solve it. 
We consider
each BS as an independent learning agent that controls the local UE transmit powers, the local IRS reflect beamformer, and its combiners via only index gradient variables. 
We design the state, action, and reward function for each BS to  capture the interdependencies among the design choices made at different BSs.
We further develop an information-sharing scheme where only limited information among neighboring BSs is exchanged to cope with the non-stationarity issue caused by the coupling between the actions at other BSs.
Through numerical simulations, we show that
our proposed scheme outperforms the conventional baselines 
for data rate maximization.

%% file: formulation.tex
\section{Multi-cell Systems with Multiple IRSs}
\label{sec:formulation}
In this section, we first introduce the signal model under consideration (Sec.~\ref{ssec:signal}). Then, we formulate the optimization and discuss the challenges associated with solving it (Sec.~\ref{sec:optimization}). 
    
\subsection{Signal Model}
\label{ssec:signal}
\begin{figure}[t]
  \includegraphics[width=.99\linewidth]{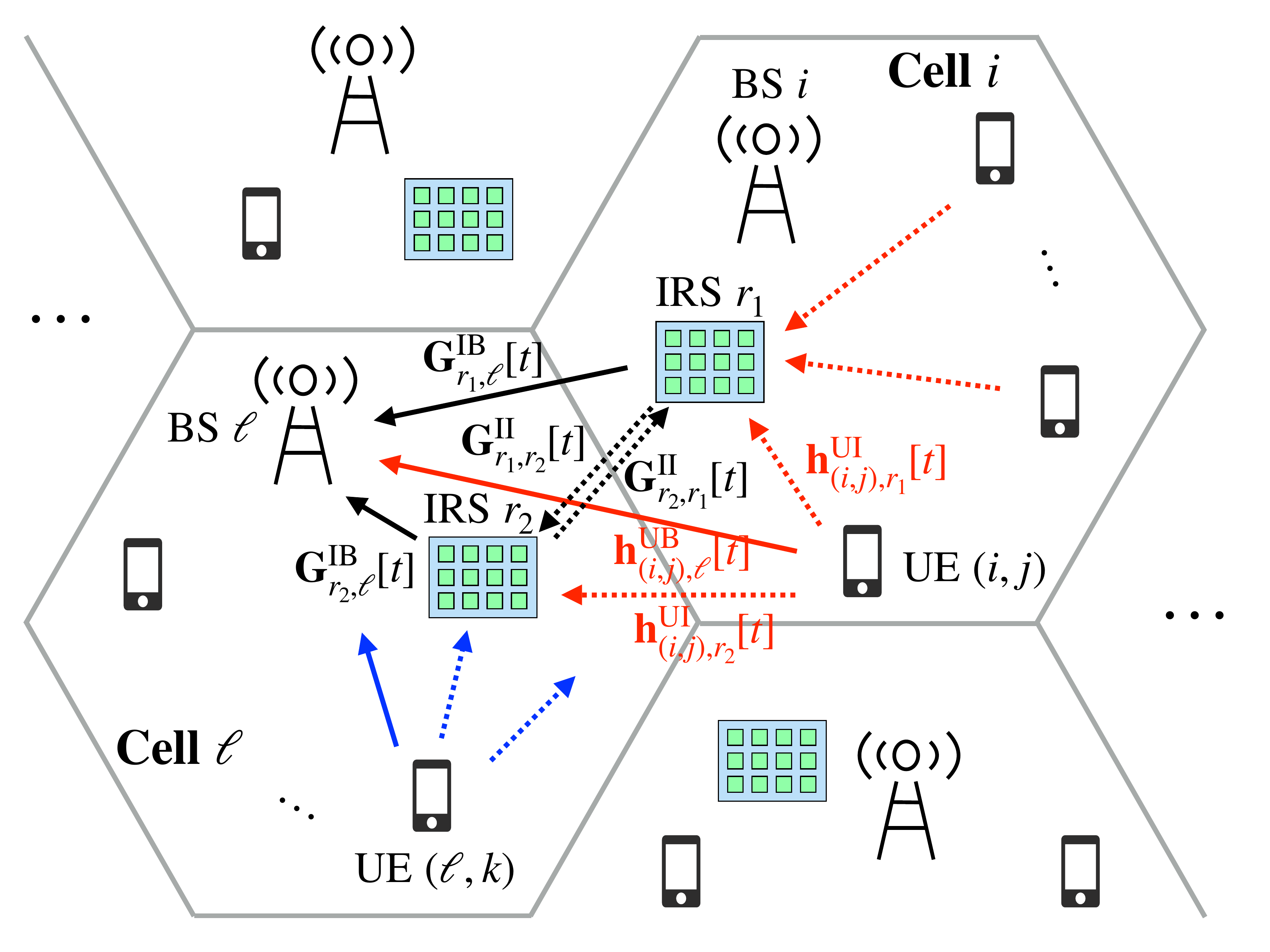}
  \centering
  \caption{Depiction of a multi-IRS-aided  multi-cell system in the UL. 
  }
  \label{fig:model}
\end{figure}

We consider a multi-cell system with multiple IRSs for the uplink (UL) as depicted in Fig.~\ref{fig:model}.
The system is comprised of a set of  $L$ cells
$\mathcal{L}= \{ 1, ..., L \}$ and $R$ IRSs
$\mathcal{R} = \{ 1, ..., R \}$. For simplicity we assume that each cell has one IRS, i.e., $R=L$, though 
our method can be readily generalized to the case where $R\neq L$. 
The IRSs are indexed such that cell $\ell$ contains IRS $\ell$.

Each cell $\ell\in\mathcal{L}$  contains
(i) $K_\ell$  UEs with single antenna, denoted by 
$\mathcal{K}_\ell = \{1, ..., K_\ell \}$, (ii) an IRS with $N_\ell$ reflecting elements, denoted by $\mathcal{N}_\ell = \{1, ..., N_\ell \}$, and (iii) a BS with $M_\ell$ antennas denoted by  $\mathcal{M}_\ell = \{ 1, ..., M_\ell\}$. 
We let  UE $(i,j)$ refer to UE $j$ in cell $i$.
The received signal vector at BS $\ell \in \mathcal{L}$ at the $t$th channel instance is given by
\begin{align}
    &{\bf y}_\ell[t] 
     = \sum_{i\in \mathcal{L}} \sum_{j \in \mathcal{K}_i} 
     \bigg\{
     \bigg(
     {\bf h}^{{\rm UB}}_{(i,j), \ell}[t] 
    +\sum_{r \in \mathcal{R}}   {\bf G}^{\rm IB}_{ r,\ell }[t] {\bf \Phi}_{r}[t] 
    {\bf h}^{{\rm UI}}_{(i,j), r}[t] 
    \nonumber
    \\
    & +\sum_{r_2 \in \mathcal{R}} \sum_{r_1 \in \mathcal{R}\setminus{\{r_2\}}}
    {\bf G}^{\rm IB}_{ r_2, \ell }[t] {\bf \Phi}_{r_2}[t] 
    {\bf G}^{\rm II}_{ r_1, r_2 }[t]
    {\bf \Phi}_{r_1}[t] 
    {\bf h}^{{\rm UI}}_{(i,j), r_1}[t] 
    \bigg)
    \nonumber \\
    &  \hspace{1.7cm}
    \sqrt{p_{i,j}[t]} s_{i,j} [t] 
    \bigg\}
    + {\bf n_\ell}[t],
    \label{eq:received}
\end{align}

\vspace{-1mm}
\noindent
where ${\bf h}^{{\rm UB}}_{(i,j), \ell}[t] \in \mathbb{C}^{M_\ell \times 1}$ is the direct channel from UE $(i,j)$ to BS $\ell$, ${\bf h}^{{\rm UI}}_{(i,j), r}[t] \in \mathbb{C}^{N_r \times 1}$ is the channel from UE $(i,j)$ to IRS $r\in \mathcal{R}$, 
${\bf G}^{\rm IB}_{r,\ell}[t] \in \mathbb{C}^{M_\ell \times N_r}$ is the channel from IRS $r$ to BS $\ell$, and 
${\bf G}^{\rm II}_{ r_1, r_2 }[t] \in \mathbb{C}^{N_{r_2} \times N_{r_1}}$ is the channel from IRS $r_1$ to IRS $r_2$, $r_1\neq r_2$. 
Also, $p_{i,j}[t] \in \mathbb{R}^+$ is the transmit power and $s_{i,j}[t] \in \mathbb{C}$ is the transmit symbol of UE $(i,j)$, where $\mathbb{E}[|s_{i,j}[t]|^2]=1$.
The noise vector ${\bf n}_{\ell}[t] \in \mathbb{C}^{M_\ell \times 1}$ at BS $\ell$ is assumed to be distributed according to zero mean complex Gaussian with covariance matrix $\sigma^2 {\bf I}$, i.e., ${\bf n}_\ell[t] \sim \mathcal{CN} ( {\bf 0}, \sigma^2 {\bf I} )$, where ${\bf I}$ denotes the identity matrix and $\sigma^2$ is the noise variance.
Finally, ${\bf \Phi}_r[t] = {\rm diag}(\phi_{r,1}[t], \phi_{r,2}[t], ..., \phi_{r,N_r}[t]) \in \mathbb{C}^{N_r \times N_r}$ is a diagonal matrix with its diagonal entries representing the beamforming vector of IRS $r \in \mathcal{R}$. 
$\phi_{r,n}[t]$, $n \in \mathcal{N}_r$, is modeled as $\phi_{r,n}[t] = a_{r,n}[t]e^{j 2\pi \theta_{r,n}[t]} \in \mathbb{C}$, incurring the signal attenuation $a_{r,n}[t] \in [0,1]$ and phase shift $\theta_{r,n}[t] \in [0,2 \pi)$.

In \eqref{eq:received}, we consider the channels with three different paths from UE $(i,j)$ to BS $\ell$: (i) the direct channel, (ii) the channel after one reflection from the IRSs (the sum over $r$), and (iii) the channel after two reflections from the IRSs (the sum over $r_1, r_2$).
Higher order reflections can also be incorporated in \eqref{eq:received}, i.e., signals reflected from more than two IRSs; we focus on up to the second-order reflections due to a large attenuation induced by multiple reflections between IRSs.

We assume that a linear combiner ${\bf z}_{\ell, k}[t] \in \mathbb{C}^{M_\ell \times 1}$ is employed at BS $\ell$ to restore $s_{\ell, k}[t]$ from ${\bf y}_{\ell}[t]$, which yields
\begin{equation}
    {\hat y}_{\ell,k}[t] = {{\bf z}^H_{\ell, k}[t]} {\bf y}_{\ell}[t],
\end{equation}
where superscript $H$ denotes the conjugate transpose.

%% file: optimization.tex
\subsection{Problem Formulation and Challenges}
\label{sec:optimization}


We aim to maximize 
the sum-rate over all the UEs in the network
through design of the
UE powers $ \{ {p}_{\ell, k}[t] \}_{\ell, k}$, BS combiners $\{ {\bf z}_{\ell, k}[t] \}_{\ell, k}$, and IRS beamformers $\{ {\boldsymbol \phi}_r[t] \}_{r}$, where ${\boldsymbol \phi}_r[t] = [\phi_{r,1}[t], \phi_{r,2}[t], ..., \phi_{r,N_r}[t]]^\top \in \mathbb{C}^{N_r \times 1}$ is the IRS beamforming vector on the diagonal of ${\boldsymbol \Phi}_r[t]$, i.e., ${\boldsymbol \Phi}_r[t] = {\rm diag}( {\boldsymbol \phi}_r[t] )$. 
With ${\rm SINR}_{\ell,k}[t]$ as the signal-to-interference ratio (SINR) of UE $(\ell,k)$, we propose
the following optimization problem:
%
\begin{align}
& \text{maximize} & & 
\sum_{\ell \in \mathcal{L}} \sum_{k \in \mathcal{K}_\ell}  \log_2(1+ {\rm SINR}_{\ell,k}[t])
\nonumber
\\
& \text{subject to}
& &  p_{\ell, k}[t] \in \mathcal{P}, \;\; { \bf z}_{\ell, k}[t]  \in \mathcal{Z}, \;\; {\boldsymbol \phi}_r[t] \in \mathcal{Q}, \;\; \forall \ell, \forall k, \forall r,
\nonumber
\\
& \text{variables} 
& & \{ {p}_{\ell, k}[t] \}_{\ell, k} , \; \{ {\bf z}_{\ell, k}[t] \}_{\ell, k} , \; \{ {\boldsymbol \phi}_{r}[t] \}_{r}, 
\label{eq:opt}
\end{align}
where $\mathcal{P}$ is the set of power values,
$\mathcal{Z}$ is 
the codebook for BS combiners,
and $\mathcal{Q}$ is the codebook for IRS beamformers.\footnote{A codebook structure can be employed for IRS because IRS is in practice controlled by a field-programmable gate array (FPGA) circuit where FPGA stores a set of coding sequences \cite{cui2014coding}.}


The problem in \eqref{eq:opt} is an optimization problem at time $t$, where $t \in \mathcal{T}= \{0, T,  2T,  ...\}$, i.e., the
optimization of variables is performed once every $T$ time instances.
If the instantaneous channels 
${\bf h}^{{\rm UB}}_{(i,j), \ell} [t]$, 
${\bf h}^{{\rm UI}}_{(i,j), r}[t]$,
${\bf G}^{\rm IB}_{ r,\ell }[t]$ and ${\bf G}^{\rm II}_{ r_1, r_2 }[t]$
in \eqref{eq:received}
are all known, then conventional optimization methods, e.g., successive convex approximation or integer programming, could be applied, since ${\rm SINR}_{\ell,k}[t]$ in \eqref{eq:opt} can be formulated as \eqref{eq:SINR} (shown at the top of the next page) with the known channels.
However, IRS-assisted wireless networks face 
the following challenges in practice:
\begin{itemize}[leftmargin=3mm]
    \item \emph{IRS channel acquisition:} Although most of the works, e.g.,~\cite{8990007,zhi2020uplink,9162705,zhang2020intelligent}, assume that channels are perfectly known, this assumption is impractical because an IRS is passive and often does not have sensing capabilities.  
    While special IRS hardware with the ability to estimate the concatenated channels does exist~\cite{wang2020channel},
    the time overhead could easily overwhelm the coherent channel resources especially when there are multiple IRSs. 
    \item \emph{Dynamic channels:} Channel dynamics in wireless environments adds another degree of difficulty to channel acquisition  and estimation. This makes solving 
    the optimization in \eqref{eq:opt} impossible with conventional model-based optimization approaches, due to dynamic and unknown channels.
    \item \emph{Centralization:} A centralized implementation to solve \eqref{eq:opt} would require gathering all the information at a central point, which is  impractical in our setting.
    Given the interdependencies among the design variables taken by different cells and their impact on the overall objective function, 
    distributed optimization of the variables
    in \eqref{eq:opt} is challenging. 
\end{itemize}

To address these challenges, we  convert \eqref{eq:opt}  into \textit{a sequential decision making problem}, where the variables are designed via successive interactions with the environment through \textit{deep reinforcement learning} (DRL).
While conventional DRL assumes a centralized implementation,
we develop a \textit{multi-agent DRL} approach, where each BS acts as an independent agent in charge of tuning its local UEs transmit powers, local IRS beamformer, and  combiners. 
To cope with the \textit{non-stationarity} issue of multi-agent DRL~\cite{marinescu2017prediction},
we carry out the learning through limited information-sharing among neighbouring BSs.

\begin{figure*}[t!]
\begin{equation}
\resizebox{.99\linewidth}{!}{$
{\rm SINR}_{\ell,k}[t] = \frac{ {p_{\ell,k}[t]} \bigg| {\bf z}^H_{\ell,k}[t] \bigg( {\bf h}^{\rm UB}_{ (\ell,k),\ell}[t] + \sum\limits_{r \in \mathcal{R}} {\bf G}^{\rm IB}_{r,\ell}[t] {\bf \Phi}_r[t] {\bf h}^{\rm UI}_{ (\ell,k),r}[t] 
+ \sum\limits_{r_2 \in \mathcal{R}} \sum\limits_{r_1 \in \mathcal{R}\setminus{\{r_2\}}}  
{\bf G}^{\rm IB}_{r_2,\ell}[t] 
{\bf \Phi}_{r_2}[t] 
{\bf G}^{\rm II}_{r_1,r_2}[t] {\bf \Phi}_{r_1}[t] 
{\bf h}^{\rm UI}_{ (\ell,k),r_1}[t]
\bigg)\bigg|^2} 
{\hspace{-4mm}
\sum\limits_{(i,j) \ne (\ell,k)}^{} {p_{i,j}[t]} \bigg| {\bf z}^H_{\ell,k}[t] \bigg( {\bf h}^{\rm UB}_{ (i,j),\ell}[t] + \sum\limits_{r \in \mathcal{R}} 
{\bf G}^{\rm IB}_{r,\ell}[t]
{\bf \Phi}_r[t] {\bf h}^{\rm UI}_{ (i,j),r}[t] 
+ \sum\limits_{r_2 \in \mathcal{R}} \sum\limits_{r_1 \in \mathcal{R}\setminus{ \{r_2\} }}  
{\bf G}^{\rm IB}_{r_2,\ell}[t]
{\bf \Phi}_{r_2}[t] {\bf G}^{\rm II}_{r_1,r_2}[t]
{\bf \Phi}_{r_1}[t] {\bf h}^{\rm UI}_{ (i,j),r_1}[t] 
\bigg)\bigg|^2 
+ \sigma^2
}$}
\label{eq:SINR}
\end{equation}
\hrulefill
\end{figure*}





%% file: MADRL.tex
\section{Multi-agent DRL Framework Design}
\label{sec:alg}

In this section, we first introduce the information collection process at the BSs and design an information-sharing scheme
(Sec.~\ref{subsec:DRL1}). We then formulate a Markov decision process (MDP) (Sec.~\ref{subsec:MDP}) and propose a dynamic control scheme 
(Sec.~\ref{subsec:MARL}) to solve our optimization from Sec.~\ref{sec:optimization}.

\subsection{Local Observations and Information Exchange}\label{subsec:DRL1}

%
We consider a setting where each BS only acquires {\it scalar effective channel powers} from {\it a subset of UEs.}
%
When UE $(i,j)$ transmits a pilot symbol with power ${p}_{i,j}[t]$, 
BS $\ell$ measures the scalar effective channel power $|\hat {h}_{(i,j),\ell,k}[t]|^2 \in \mathbb{R}$ (after combining with ${\bf z}_{\ell, k}[t]$), $k \in \mathcal{K}_\ell$, which is given by
\begin{equation}
    |\hat {h}_{(i,j),\ell,k}[t]|^2 = |{\bf z}_{\ell, k}^H[t] \hat {\bf h}_{(i,j),\ell}[t]|^2,
    \label{eq:h_scalar}
\end{equation}
%
where $\hat {h}_{(i,j),\ell,k}[t] \in \mathbb{C}$ is the {\it scalar effective channel}.
The vector $\hat {\bf h}_{(i,j),\ell}[t] \in \mathbb{C}^{M_\ell \times 1}$ is the \textit{effective channel} from UE $(i,j)$ to BS $\ell$ (before combining), which is expressed as follows:
\begin{multline}
    \hat {\bf h}_{(i,j),\ell}[t] = \sqrt{p_{i,j}[t]} \bigg( {\bf h}^{\rm UB}_{ (i,j),\ell}[t]
     + \sum_{r \in \mathcal{R}}{\bf G}^{\rm IB}_{r,\ell}[t]
     {\bf \Phi}_r[t] {\bf h}^{\rm UI}_{ (i,j),r}[t] 
    \\
    + \sum_{r_2 \in \mathcal{R}} \sum_{r_1 \in \mathcal{R}\setminus\{r_2\}}  {\bf G}^{\rm IB}_{r_2,\ell}[t]
    {\bf \Phi}_{r_2}[t] 
    {\bf G}^{\rm II}_{r_1,r_2}[t]
    {\bf \Phi}_{r_1}[t] {\bf h}^{\rm UI}_{ (i,j),r_1}[t] 
    \bigg).
    \label{eq:h_vector}
\end{multline}

BS $\ell$ collects the scalar effective channel powers of the links (i) from local UEs (in cell $\ell$) to BS $\ell$, (ii)  from neighbouring UEs (not in cell $\ell$) to BS $\ell$, and (iii) from local UEs to neighbouring BSs.
BS $\ell$ measures (i) and (ii) as local observations, but needs to receive (iii), which cannot be measured by BS $\ell$, from neighbouring BSs.
Additionally, BS $\ell$ receives a \textit{penalty value} from neighbouring BSs, where the penalty value is used for designing the reward function and will be formalized in Sec.~\ref{subsubsec:reward}.
Note that concurrent estimation of the scalar effective channel powers of multiple UEs can be performed by 
UE-specific reference signals in the Long-Term Evolution (LTE) standard \cite{LTEstandard}. Acquiring scalar effective channel powers from only a subset of UEs 
lowers
the CSI acquisition overhead compared to the conventional method of acquiring large-dimensional vector or matrix CSI from individual UE for each IRS.


To clarify which neighbouring UEs are included in (ii) and which neighbouring BSs are included in (iii), we define two sets of cell indices.
First, we define the set of indices of {\it dominantly interfering} neighboring cells, $\mathcal{B}^{(1)}_{\ell}[t]$. 
UEs in cell $i \in \mathcal{B}^{(1)}_{\ell}[t]$ are dominantly interfering with the data link of local UEs (in cell $\ell$). Formally, 
$  \forall i\in \mathcal{B}^{(1)}_{\ell}[t],~ \forall i'\in \mathcal{L}\backslash \mathcal{B}^{(1)}_{\ell}[t] \backslash{\{\ell\}}$, 
we have
\begin{equation}
    \sum_{j \in \mathcal{K}_i} \| \hat {\bf h}_{(i,j),\ell}[t] \|_2^2 \ge  \sum_{j \in \mathcal{K}_{i'}} \| \hat {\bf h}_{(i',j),\ell}[t] \|_2^2.
\end{equation} 
The size of this set is a control variable  $B^{(1)}=|\mathcal{B}^{(1)}_{\ell}[t]|$.
For (ii), then, we include  neighbouring UEs in cell $i \in \mathcal{B}^{(1)}_{\ell}[t]$.

Second, we  define the set of indices of {\it dominantly interfered} neighboring cells, $\mathcal{B}^{(2)}_{\ell}[t]$. The data links of UEs in cell $i \in \mathcal{B}^{(2)}_{\ell}[t]$ are dominantly interfered by local UEs (in cell $\ell$). Formally,
$\forall i\in \mathcal{B}^{(2)}_{\ell}[t],~ \forall i'\in \mathcal{L}\backslash \mathcal{B}^{(2)}_{\ell}[t] \backslash{\{\ell\}}$, we have 
\begin{equation}
    \sum_{k \in \mathcal{K}_\ell} \| \hat {\bf h}_{(\ell,k),i}[t] \|_2^2 \ge  \sum_{k \in \mathcal{K}_{\ell}} \| \hat {\bf h}_{(\ell,k),i'}[t] \|_2^2.
\end{equation}
The size of this set is a control variable $B^{(2)}=|\mathcal{B}^{(2)}_{\ell}[t]|$.
For (iii), then, we include neighbouring BSs of cell $i \in \mathcal{B}^{(2)}_{\ell}[t]$.

The effective channel gain, used in defining $\mathcal{B}^{(1)}_{\ell}[t]$ and $\mathcal{B}^{(2)}_{\ell}[t]$,
can be acquired by the antenna circuit before digital processing (e.g., from the automatic gain control (AGC) circuit \cite{mo2017channel}), without 
the explicit effective channel vector or combiner implementation.
BS $\ell$ also measures ${ {\rm SINR}}_{\ell, k}[t]$  of all local UEs, by measuring the received signal strength indicator (RSSI) 
and the reference signal received power (RSRP), which are the conventional measures to evaluate the signal quality in LTE standards \cite{LTEstandard}.
Using the SINRs, BS $\ell$ then calculates the achievable data rate
of UE $(\ell,k)$ as
${R}_{\ell, k}[t] =
{
\log_2 (1+ { {\rm SINR}}_{\ell, k}[t])
}$.
Here, we omit the bandwidth parameter, assuming the same bandwidth for all the data links.

\subsection{Markov Decision Process Model}
\label{subsec:MDP}
We formulate the decision making process of each BS as an MDP with states, actions, and rewards:
\vspace{0.5mm}
\subsubsection{State}
\label{subsubsec:state}
We define the state space of BS $\ell$ as 
\begin{equation}
    \mathcal{S}_{\ell}[t] = 
\mathcal{S}_{\ell,1}[t] \bigcup \mathcal{S}_{\ell,2}[t] \bigcup
\mathcal{S}_{\ell,3}[t] \bigcup
\mathcal{S}_{\ell,4}[t],
\end{equation}
where each constituent set is described below. 

\textbf{(i) Local channel information.} 
   $\mathcal{S}_{\ell,1}[t]$  consists of the scalar effective channel powers from local UEs observed at two consecutive times $t-T$ and $t$, given by
%
    \begin{equation}
   \mathcal{S}_{\ell,1}[t] = \{ |\hat {h}_{(\ell,j),\ell,k}[t-T]|^2, \; |\tilde {h}_{(\ell,j),\ell,k}[t]|^2 \}_{j \in \mathcal{K}_\ell, k \in \mathcal{K}_\ell }.
   \nonumber
   \end{equation}
   %
   Here, $|\hat {h}_{(\ell,j),\ell,k}[t-T]|^2$ can be obtained from \eqref{eq:h_scalar} at time $t-T$, and $|\tilde {h}_{(\ell,j),\ell,k}[t]|^2$ is a version of \eqref{eq:h_scalar} obtained at time $t$ using previous-time variables $p_{\ell,k}[t-T]$, ${\bf z}_{\ell,k}[t-T]$, and $ \{ \boldsymbol{\phi}_{r}[t-T] \}_{r \in \mathcal{R}}$. 
    Having them enables us to capture the effect of channel variation over time.

 \textbf{(ii) From-neighbor channel information.} 
    $\mathcal{S}_{\ell,2}[t]$ contains the scalar effective channel powers from UE $(i,j)$ in neighboring cell $i$, and the index $i$, for $i \in \mathcal{B}^{(1)}_{\ell}[t]$, 
   $j \in \mathcal{K}_i$. Formally,
   \begin{equation}
   \mathcal{S}_{\ell,2}[t] = \{|\hat {h}_{(i,j),\ell,k}[t-T]|^2 \}_{j \in \mathcal{K}_i, k \in \mathcal{K}_\ell, i \in \mathcal{B}^{(1)}_{\ell}[t]} \bigcup \{i\}_{i \in \mathcal{B}^{(1)}_{\ell}[t]}.
   \nonumber
   \end{equation}
   %
This set captures the interference from neighbor UEs to cell $\ell$. 
   

\textbf{(iii) To-neighbor channel information.} 
 $\mathcal{S}_{\ell,3}[t]$ contains the scalar effective channel powers from local UE $(\ell,k)$ to BS $i$, and the index $i$,  for $i \in \mathcal{B}^{(2)}_{\ell}[t]$, $k \in \mathcal{K}_\ell$. Formally,
 \begin{equation}
     \mathcal{S}_{\ell,3}[t] =  \{ |\hat {h}_{(\ell,k),i,j}[t-T]|^2\}_{j \in \mathcal{K}_i ,k \in \mathcal{K}_\ell, i \in \mathcal{B}^{(2)}_{\ell}[t]} \bigcup \{ i \}_{i \in \mathcal{B}^{(2)}_{\ell}[t]}.
     \nonumber
 \end{equation}
 %
This set captures the amount of interference that local UEs in cell $\ell$ inflict on neighboring cells. 
This information enables BS $\ell$ to adjust the transmit powers of local UEs to reduce interference to the neighboring cells.

\textbf{(iv) Previous local variables and local sum-rate.}
    $\mathcal{S}_{\ell,4}[t]$ consists of previous local variables, i.e., $\{ {p}_{\ell, k}[t-T] \}_{k \in \mathcal{K}_\ell}$, $\{ {\bf z}_{\ell, k}[t-T] \}_{k \in \mathcal{K}_\ell}$, and  ${\boldsymbol \phi}_{\ell}[t-T] $, and the local sum-rate $R_\ell[t-T] = \sum_{k\in \mathcal{K}_{\ell}} {R}_{\ell,k}[t-T]$. Formally,
    \begin{equation}
        \mathcal{S}_{\ell,4}[t] = \{ {p}_{\ell, k}[t-T], {\bf z}_{\ell, k}[t-T] \}_{k \in \mathcal{K}_\ell} 
    \bigcup \{ {\boldsymbol \phi}_{\ell}[t-T], R_\ell[t-T]\}.
    \nonumber
    \end{equation}
    
\subsubsection{Action}
The action space is defined as
\begin{equation}
    \mathcal{A}_\ell[t] = \{ b^{p}_{\ell,1}[t], ..., b^{p}_{\ell,K_\ell}[t], b^{z}_{\ell,1}[t], ..., b^{z}_{\ell,K_\ell}[t], b^{\phi}_{\ell}[t] \},
    \label{eq:action}
\end{equation}
where $b^{p}_{\ell,k}[t]$, $b^{z}_{\ell,k}[t]$, $b^{\phi}_{\ell}[t]$ are the \textit{index gradient variables} used for updating the local UE $(\ell, k)$ transmit power, combiner $k$ of BS $\ell$, and local IRS $\ell$ reflect beamformer. These index gradient variables are defined over a binary $\{ -1,1 \}$, or ternary $\{ -1, 0, 1 \}$ alphabet as we will describe in  Sec.~\ref{sec:sim}. 



Once BS $\ell$ determines the action in \eqref{eq:action}, the BS feeds forward $b^{p}_{\ell,k}[t]$ to UE $(\ell,k)$, which then updates its power index as
\begin{equation}
    i^{p}_{\ell,k}[t] = i^{p}_{\ell,k}[t-T] + b^{p}_{\ell,k}[t].
\end{equation}
The power of UE $(\ell,k)$ is set to $p_{\ell,k}[t] = \mathcal{P}(i^{p}_{\ell,k}[t])$, $k \in \mathcal{K}_\ell$, where $\mathcal{P}(i)$ denotes $i$-th element of the power set $\mathcal{P}$ in \eqref{eq:opt}. 
The BS also feeds forward  $b^{\phi}_{\ell}[t]$ to IRS $\ell$, which then updates its beamformer index as
\begin{equation}
    i^{\phi}_{\ell}[t] = i^{\phi}_{\ell}[t-T] + b^{\phi}_{\ell}[t].
\end{equation} 
The beamformer of IRS $\ell$ is set to ${\boldsymbol \phi}_\ell[t] = \mathcal{Q}(i^{\phi}_{\ell}[t])$ where $\mathcal{Q}(i)$ is the $i$-th vector in the codebook $\mathcal{Q}$ in \eqref{eq:opt}.
Finally, the combiner index  is updated as 
\begin{equation}
    i^{z}_{\ell,k}[t] = i^{z}_{\ell,k}[t-T] + b^{z}_{\ell,k}[t].
\end{equation}
The combiner $k $ of BS $\ell$ is set to ${\bf z}_{\ell,k}[t] = \mathcal{Z}(i^{z}_{\ell,k}[t])$.

\vspace{.5mm}
\subsubsection{Reward}
\label{subsubsec:reward}

Aiming to only maximize the local sum-rate at each BS
could increase the interference to the neighboring cells. 
To incorporate the entire system performance, we design the reward $r_{\ell}[t]$ including penalty terms as 
%
\begin{equation}
    r_{\ell}[t] = \sum_{k \in \mathcal{K}_\ell} { R}_{\ell,k}[t] - \sum_{i \in \mathcal{B}^{(2)}_{\ell}[t] } P_{\ell,i}[t],
\end{equation}
%
where the first term is the sum-rate of cell $\ell$ and the second term is the sum of penalties. 
The penalty $P_{\ell,i}[t]$ is the rate loss of the dominantly interfered cell $i$ caused by the interference of local UEs (in cell $\ell$), which is calculated at BS $i$ as
\begin{align}
    &\hspace{-2mm} P_{\ell,i}[t] = \sum_{j \in \mathcal{K}_i} P_{\ell,(i,j)}[t]=   \sum_{j \in \mathcal{K}_i} \bigg[  - { R}_{i,j}[t]
    \nonumber
    \\
    &\hspace{-2mm} + \log_2 \bigg(1+ 
    \frac{ |\hat {h}_{(i,j),i,j}[t]|^2 }
    { \sum\nolimits_{(i',j') \ne (i,j), i' \ne \ell } |\hat {h}_{(i',j'),i,j}[t]|^2  + \sigma^2}
    \bigg) 
     \bigg],
     \label{eq:penalty}
\end{align}
%
where $P_{\ell,(i,j)}[t]$ is the rate loss of UE $(i,j)$ caused by the interference of local UEs in cell $\ell$.
A similar reward function was found to be effective for multi-agent DRL-based beamforming~\cite{ge2020deep}.
The $\log(\cdot)$ term in \eqref{eq:penalty} denotes the data rate of UE $(i,j)$ without the interference of the local UEs in cell $\ell$,
while $R_{i,j}[t]$ is the data rate including the interference.
If there is no interference, 
the two terms cancel with each other, leading to zero penalty. Otherwise, the penalty is positive.

 \begin{algorithm}[h]
 \caption{Dynamic control based on multi-agent DRL.}
 \label{al:drl}
 {\footnotesize
 \begin{algorithmic}[1]
 \small
  \STATE Establish a train DQN with random weights ${\bf w}_\ell$, a target DQN with random weights ${\bf w}_\ell^-$, an empty experience pool $\mathcal{Y}_\ell$ with $|\mathcal{Y}_\ell|=0$, and a pool size $M^{\rm pool}_{\ell}$.
  Set the discount factor $\gamma_\ell$, initial $\epsilon$-greedy value $\epsilon_\ell(0)$, mini-batch size $M^{\rm batch}_\ell$, and DQN-aligning period $T_{\rm align}$, $\forall \ell \in \mathcal{L}$.
  \STATE Agent $\ell$ (BS $\ell$) randomly initializes the design variables $\{p_{\ell,k}[0]\}_{k \in \mathcal{K}_\ell}$, $\{{\bf z}_{\ell,k}[0]\}_{k \in \mathcal{K}_\ell}$, and $\boldsymbol{\phi}_{\ell}[0]$, and informs local UEs and local IRS of the initial variables, $\forall \ell \in \mathcal{L}$. 
  \STATE Agent $\ell$ selects its action $a_\ell \in \mathcal{A}$ randomly and executes it, $\forall \ell \in \mathcal{L}$.
    \STATE $t \leftarrow T$.    Agent $\ell$ observes the next state $s_\ell'$, $\forall \ell \in \mathcal{L}$.
  \REPEAT
    \STATE { $s_\ell \leftarrow s'_\ell$ }
    \STATE { Agent $\ell$ selects its action $a_\ell$ at time $t$ based on $\epsilon$-greedy policy, $\forall \ell \in \mathcal{L}$: With probability $\epsilon_\ell(t)$, agent $\ell$ selects random action $a_\ell$, and with probability $1-\epsilon_\ell(t)$, agent $\ell$ selects $a_\ell = {\arg\max}_{a \in \mathcal{A}} Q(s_\ell,a, {\bf w}_\ell)$. 
    }
    \STATE Agent $\ell$ executes its action, $\forall \ell \in \mathcal{L}$.
    \STATE $t \leftarrow t+T$. Agent $\ell$ observes the next state $s'_\ell$ and gets the reward $r_\ell$, $\forall \ell \in \mathcal{L}$.
    \STATE Agent $\ell$ stores the new experience $< s_\ell, a_\ell, r_\ell, s'_\ell >$ in its own experience pool $\mathcal{Y}_\ell$, $\forall \ell \in \mathcal{L}$.
    \IF{$|\mathcal{Y}_\ell| \ge M^{\rm batch}_\ell $}
    \STATE Agent $\ell$ samples a mini-batch consisting of $M^{\rm batch}_\ell$ experiences from its experience pool $\mathcal{Y}_\ell$, $\forall \ell \in \mathcal{L}$.
    \STATE Agent $\ell$ updates the weights ${\bf w}_\ell$ of its train DQN using back propagation, $\forall \ell \in \mathcal{L}$.
    \STATE Agent $\ell$ updates the weights of its target DQN ${\bf w}_\ell^- \leftarrow {\bf w}_\ell$ every $T_{\rm align}$, $\forall \ell \in \mathcal{L}$.
    \ENDIF
  \UNTIL {Process terminates}
 \end{algorithmic}
 }
 \end{algorithm}

\subsection{Dynamic Control Scheme based on Multi-agent DRL}
\label{subsec:MARL}
In the proposed MDP, the channel values used as states are continuous variables, which 
makes conventional RL, i.e., Q-learning based on Q-table, not applicable.
We thus adopt deep Q-networks (DQN) \cite{mnih2015human}.
BS $\ell$ possesses its own train DQN, $Q(s,a,{\bf w}_\ell)$, with weights ${\bf w}_\ell$, and target DQN, $Q(s,a,{\bf w}_\ell^-)$, with weights ${\bf w}_\ell^-$, where the state $s \in \mathcal{S}$ and action $a \in \mathcal{A}$ are defined in Sec. \ref{subsec:MDP}.
The pseudocode of the proposed dynamic control scheme based on multi-agent DRL is provided in Algorithm~\ref{al:drl}.
Our algorithm follows a decentralized training with decentralized execution (DTDE) framework, where both training and execution are independently carried out at each agent. Therefore, our algorithm is independent of the UEs in other surrounding agents (BSs). Further, our algorithm incorporates the {\it index gradient} approach for codebook-based BS combining and IRS beamforming, which is independent of the number of antennas/elements and the size of the codebook.




%% file: sim.tex
\section{Numerical Evaluation and Discussion}
\label{sec:sim}

In this section, we first describe the simulation setup (Sec.~\ref{sec:sim1}) and evaluation scenarios (Sec.~\ref{sec:sim2}). Then, we present and discuss the results~(Sec.~\ref{sec:sim3}).  
\subsection{Simulation Setup}\label{sec:sim1}
\subsubsection{Parameter settings}
\begin{figure}[t]
  \centering
  \includegraphics[width=.65\linewidth]{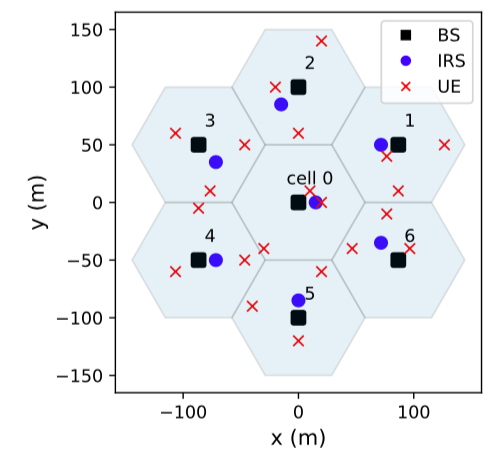}
  \caption{The cellular network with $L=7$ hexagonal cells and 100 m distance between adjacent BSs used in our simulations.}
  \label{fig:cell_plot}
\end{figure}

We consider a cellular network with $L=7$ hexagonal cells, as shown in Fig.~\ref{fig:cell_plot}.
We assume $K_\ell = 3$, $M_\ell = 5$, and $N_{r}=5$, $\forall \ell,r$, 
similar to \cite{pan2020multicell}.
The BSs are located at the center of  each cell with 10 m height, and the distance between adjacent BSs is 100 m. Each IRS is deployed nearby the BS, and UEs are randomly placed in the cells.
The set $\mathcal{P}$ for UE power control is given by $\mathcal{P} = \{ p_{\min}, p_{\min}e^{\Delta_{\rm p}}, p_{\min}e^{2\Delta_{\rm p}}, ..., p_{\max}  \}$, where $p_{\min} = 10$ dBm and $p_{\max} = 30$ dBm are the minimum and maximum transmit powers, and $\Delta_{\rm p} = (\log p_{\max} -\log p_{\min})/(|\mathcal{P}|-1)$.
For BS combiner and IRS beamformer codebooks,
we use a random vector quantization (RVQ)~\cite{au2007performance} codebook with size $|\mathcal{Z}|=|\mathcal{Q}|=30$. 
We set $\sigma^2 = -114$ dBm, $B^{(1)} = B^{(2)} =2$.


\subsubsection{Channel modeling}

We consider a single frequency band with flat fading and adopt a temporally correlated block fading channel model.
Following a common cellular standard \cite{ieeep80216}, we assume coherence time $T=5$ ms and center frequency $f_c = 2.5$ GHz. 
%
The channel vector ${\bf h}^{{\rm BS}}_{(i,j), \ell} [t]$ is modeled as
%
\begin{equation}
    {\bf h}^{{\rm UB}}_{(i,j), \ell} [t] = \sqrt{\beta^{\rm UB}_{(i,j),\ell}} {\bf u}^{{\rm UB}}_{(i,j), \ell}[t],
\end{equation}
%
where
$\beta^{\rm UB}_{(i,j),\ell} $ denotes the large-scale fading coefficient from UE $(i,j)$ to BS $\ell$,
 modeled as
\begin{equation}
    \beta^{\rm UB}_{(i,j),\ell} = \beta_0 - 10 \alpha^{\rm UB}_{(i,j),\ell} \log_{10} (d^{\rm UB}_{(i,j),\ell}/d_0).
\end{equation} 
%
Here, $\beta_0$ is the path-loss at the reference distance $d_0$, $d^{\rm UB}_{(i,j),\ell}$ is the distance between UE $(i,j)$ and BS $\ell$, and $\alpha^{\rm UB}_{(i,j),\ell}$ is the path-loss exponent between them.  
We set $\beta_0 = -30$ dB and $d_0 = 1$ m. 
${\bf u}^{\rm UB}_{(i,j), \ell}[t]$ denotes the Rayleigh fading vector, 
modeled by a first-order Gauss-Markov process \cite{sklar2001digital}: 
\begin{equation}
    {\bf u}^{{\rm UB}}_{(i,j), \ell}[t] = \rho^{{\rm UB}}_{(i,j), \ell} {\bf u}^{{\rm UB}}_{(i,j), \ell}[t-T] + \sqrt{1- ({\rho^{{\rm UB}}_{(i,j), \ell}})^2  } {\bf n}^{{\rm UB}}_{(i,j), \ell}[t], 
\end{equation}
%
where
${\bf n}^{\rm UB}_{(i,j), \ell}[t] \in \mathbb{C}^{M_{\ell} \times 1}$,  ${\bf n}^{\rm UB}_{(i,j), \ell}[t] \sim \mathcal{CN} ( {\bf 0}, {\bf I} )$, and ${\bf u}^{\rm UB}_{(i,j), \ell}[0] \sim \mathcal{CN} ( {\bf 0}, {\bf I} ) $.
The time correlation coefficient obeys the Jakes model \cite{sklar2001digital}, i.e., $\rho^{{\rm UB}}_{(i,j), \ell} = J_0(2\pi {\tilde f}^{\rm UB}_{(i,j)} T)$,
where $J_0(\cdot)$ is the zeroth order Bessel function of the first kind, and ${\tilde f}^{\rm UB}_{(i,j),\ell} = v^{\rm UB}_{(i,j),\ell} f_{\rm c}/c$ is the maximum Doppler frequency, with velocity $v^{\rm UB}_{(i,j),\ell}$ of UE $(i,j)$  and $c=3 \times 10^8$ m/s.
The same modeling for ${\bf h}^{{\rm UB}}_{(i,j), \ell} [t]$ is applied for the channels between the UEs and the IRSs, i.e., ${\bf h}^{{\rm UI}}_{(i,j), r} [t]$, $\forall i,j,r$, with path-loss exponent $\alpha^{\rm UI}_{(i,j),r}$.
%
Since IRSs are placed at the desired locations to have less variations of IRS-BS/IRS-IRS channels as compared to UE-BS/UE-IRS channels~\cite{pan2020multicell},
${\bf G}^{\rm IB}_{ r,\ell }[t]$ and ${\bf G}^{\rm II}_{ r_1, r_2 }[t]$ are assumed to be stationary.
Each entry for the channels is distributed according to {\small $ \mathcal{CN}(0,\beta^{\rm IB}_{r,\ell})$} and {\small $ \mathcal{CN}(0,\beta^{\rm II}_{r_1, r_2})$}, respectively.
 $\beta^{\rm IB}_{r,\ell}$ and $\beta^{\rm II}_{r_1, r_2}$
denote  the  large-scale  fading  coefficients with path loss exponents $\alpha^{\rm IB}_{r,\ell}$ and $\alpha^{\rm II}_{r_1, r_2}$, respectively.

%

We assume $\alpha^{\rm UB}_{(i,j),\ell} = \alpha^{\rm UB}$, $\forall i,j,\ell$, $\alpha^{\rm UI}_{(i,j),r} = \alpha^{\rm UI}$, $\forall i,j,r$, $\alpha^{\rm IB}_{r,\ell} = \alpha^{\rm IB}$, $\forall r,\ell$, and $\alpha^{\rm II}_{r_1, r_2} = \alpha^{\rm II}$, $\forall r_1,r_2$.
To model the presence of extensive obstacles and scatterers, the path-loss exponent between the UEs and BS is taken to be $\alpha^{\rm UB} = 3.75$. 
Because the IRS-aided link can have less path loss than that of direct UE-BS channel by properly choosing the location of the IRS, we set the path-loss exponents of the UE-IRS link, of the IRS-BS link, and of the IRS-IRS link to $\alpha^{\rm UI} = 2.2$,  $\alpha^{\rm IB} = 1$, and $\alpha^{\rm II} = 2$, respectively~\cite{pan2020multicell}. 
We assume $\rho^{\rm UB}_{(i,j),\ell} = \rho^{\rm UI}_{(i,j),r} =\rho$, $\forall i,j,\ell,r$ and adopt $\rho = 0.999$ ($v \approx 1$ km/h), $0.99$ ($v \approx 3$ km/h), and $0.9$ ($v \approx 9$ km/h), where $v$ is the UE speed.

\subsection{Evaluation Scenarios}\label{sec:sim2}

\subsubsection{Scenario 1. The effective channels from local UEs are not known}
In this scenario, each BS measures the scalar effective channel powers directly from received signals without explicitly obtaining the effective channels as a vector form in \eqref{eq:h_vector}.
We introduce two baselines in this scenario: RRR=(random, random, random) and MRR=(maximum, random, random). The name of each baseline is indicating how it selects its (UE power, IRS beamformer, BS combiner) variables as a tuple.
%
%
We propose DQN1, where the action space consists of $2K+1$ elements for $K$ UE powers, the IRS beamformer, and $K$ BS combiners. The index gradient variables are binary, i.e., $\{-1,1\}$. 

\subsubsection{Scenario 2. The effective channels from local UEs are known} In this scenario, each BS measures the effective channels from local UEs as the vector form in \eqref{eq:h_vector}.
Each BS is assumed to adopt a maximum ratio combiner (MRC) by finding the index as
\begin{equation}
    i^\star = {\arg\max}_{i} | \mathcal{Z}(i)^H {\bf h}[t]|^2,
\end{equation}
where ${\bf h}[t]$ is the effective channel from local UE.
We introduce several baselines: MRM=(maximum, random, MRC), FRM=(25\% of maximum, random, MRC), RRM=(random, random, MRC), and MM with no IRS=(maximum, N/A, MRC). 
MM with no IRS assumes the IRSs to be turned off.
In this scenario, we propose DQN2 and DQN3.
In DQN2, the action space consists of $K+1$ elements for $K$ UE powers and the IRS beamformer (the action space does not have the elements $b^{z}_{\ell,k}[t]$, $\forall k$ in \eqref{eq:action}). The BS combiner is designed as MRC and the index gradient variable is binary, i.e., $\{-1,1\}$.
The action space is DQN3 is the same as DQN2, except it uses a tenary index gradient variable, i.e., $\{-1,0,1\}$.

In both scenarios, the DQNs\footnote{All DQNs establish the same state space and reward function given in Sec.~\ref{subsec:MDP}. For the state information group (iv) in Sec.~\ref{subsubsec:state}, the indices of previous local variables are stored in the state.} 
are composed of an input layer, an output layer, and two fully-connected hidden layers. The input size is $6K^2 + 2K+6 = 66$. The output size is $2^{2K+1}=128$, $2^{K+1}=16$, and $3^{K+1}=81$ for DQN1, DQN2, and DQN3, respectively.
For DQN1, the number of neurons in the two hidden layers is 70 and 100; for DQN2, 40 and 30; and for DQN3, 70 and 70. The rectified linear unit (ReLU) activation function is employed.
In Algorithm~\ref{al:drl}, we adopt the $\epsilon$-greedy method with $\epsilon_\ell(t) = \max \{ \epsilon_{\min}, (1-10^{-3.5}) \epsilon_\ell (t-T) \}$, where $\epsilon_\ell(0)=0.6$ and $\epsilon_{\min} = 0.005$, $\forall \ell$. 
We consider $M^{\rm batch}_\ell = 10$, $M^{\rm pool}_\ell =300$, and $\gamma_\ell=0.7$, $\forall \ell$.
We set $T_{\rm align} = 50T$, i.e., the target DQN is updated with the weights of train DQN after a time of $50T$. We employ the RMSProp optimizer for training. 



\begin{figure}[t!]
\centering
%
\begin{subfigure}{.9\linewidth}
  \centering
  \includegraphics[width=\linewidth]{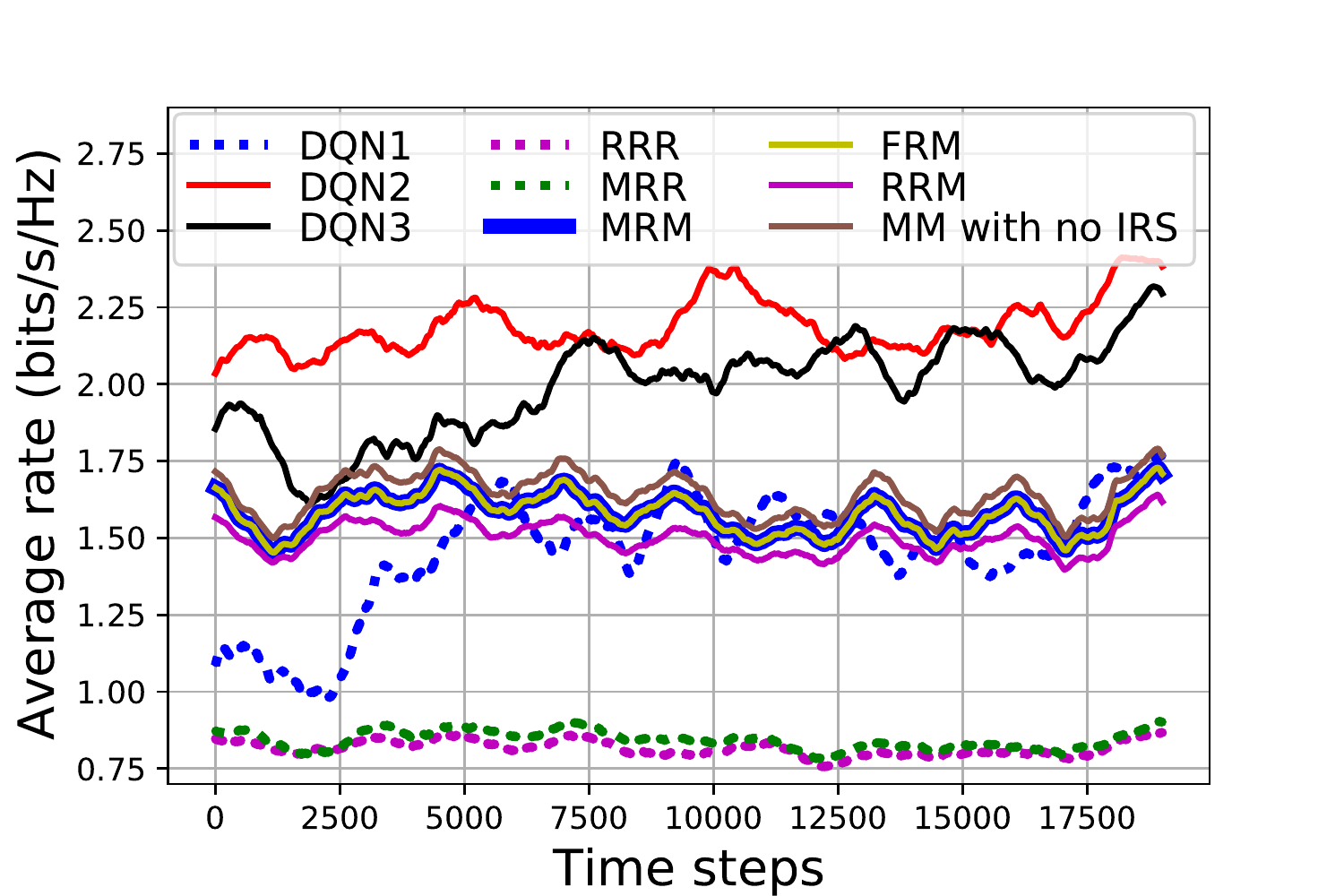}
  \caption{$\rho = 0.999$}
  \label{fig:avgrate0999}
\end{subfigure}
%
\begin{subfigure}{.9\linewidth}
  \centering
  \includegraphics[width=\linewidth]{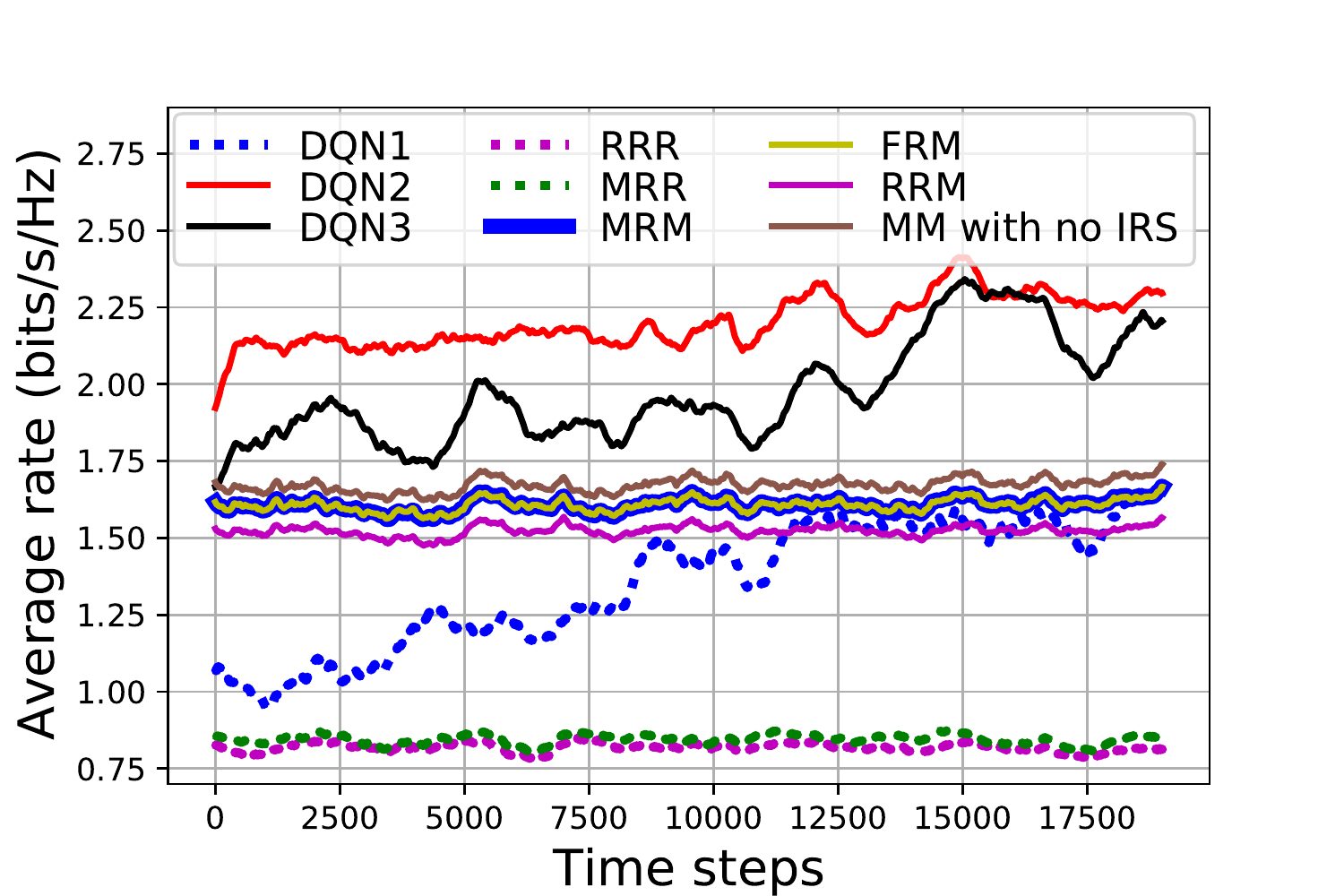}
  \caption{$\rho = 0.99$}
  \label{fig:avgrate099}
\end{subfigure}
%
\begin{subfigure}{.9\linewidth}
 \centering
 \includegraphics[width=\linewidth]{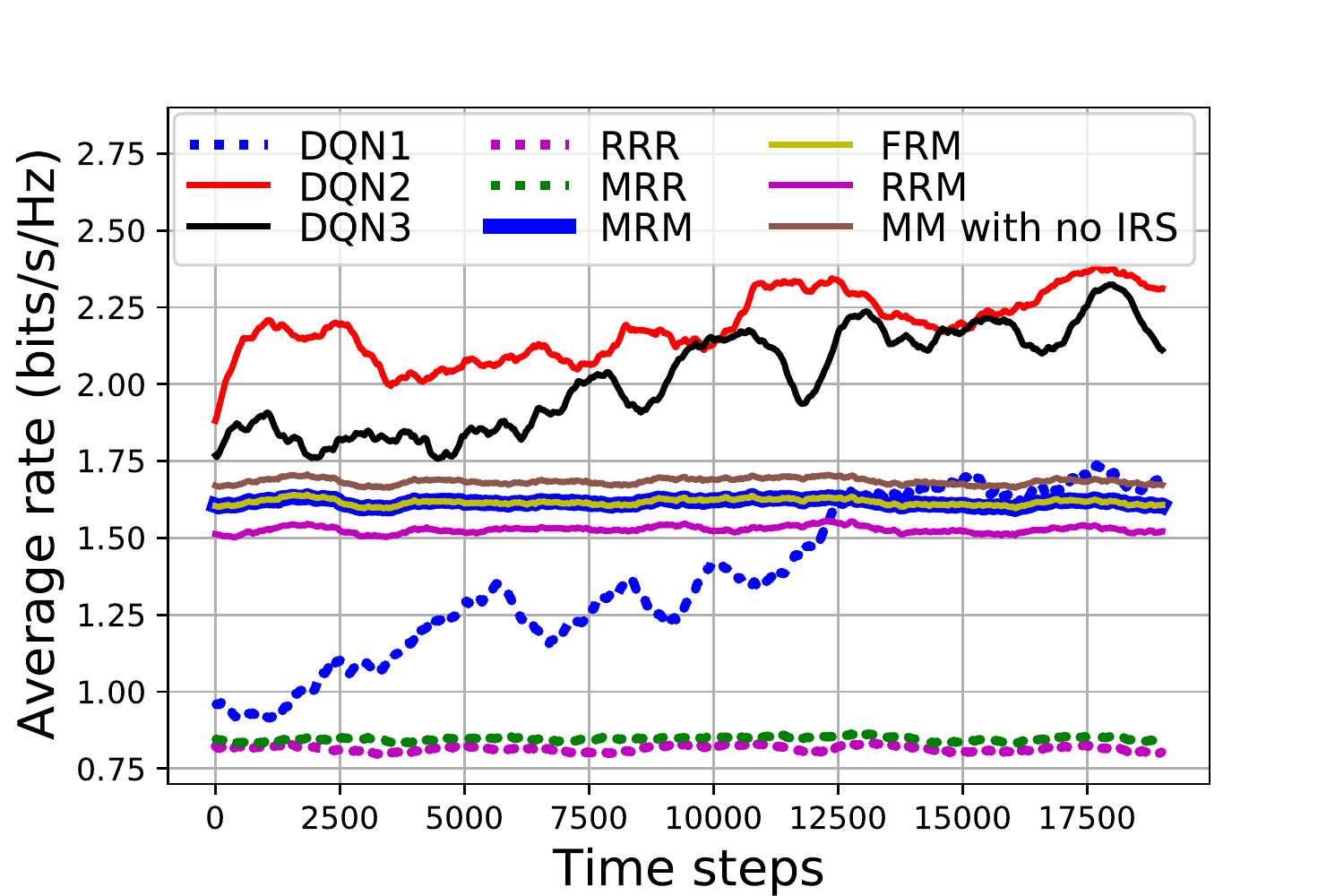}
 \caption{$\rho = 0.9$}
 \label{fig:avgrate09}
\end{subfigure}
%
\caption{
Average achievable data rates over all $21$ UEs obtained by each method with different values of $\rho$: $\rho=0.999$ in (a), $\rho=0.99$ in (b), and $\rho=0.9$ in (c). The dotted-lines and solid lines show the performance of schemes in Scenario 1 and Scenario 2, respectively.
Each data point in the plots is a moving average over the previous 1000 time slots.}
\label{fig:sim}
\end{figure}


\subsection{Simulation Results and Discussion}\label{sec:sim3}

Fig.~\ref{fig:sim} depicts the average achievable data rate over all 21 UEs with different values of time correlation coefficient $\rho$: $\rho=0.999$ in (\subref{fig:avgrate0999}), $\rho=0.99$ in (\subref{fig:avgrate099}), and $\rho=0.9$ in (\subref{fig:avgrate09}).
The dotted lines show the performance of the schemes in Scenario 1.
With varying channels, RRR and MRR select random or fixed indices for variables, and therefore have low average data rates over time.
On the other hand, DQN1 learns and adapts to the varying channels over time by exploiting the local observations and information-sharing in our sequential decision making. 

The solid lines represent the performances of schemes in Scenario 2.
The MM with no IRS gives better performance than the baselines using IRS, implying that random IRS beamforming is worse than not deploying it at all.
This also reveals the vulnerability of IRS-assisted systems to adversarial IRS
utilization.
Our DQN2 and DQN3 methods outperform the baselines, which emphasizes the benefit of carefully optimizing the IRS configuration with the rest of the cellular network.
DQN2 yields slightly better performance and converges faster than DQN3:
the faster convergence is due to neural networks training faster with a smaller number of outputs, and the better overall performance is  consistent with the observation \cite{mnih2015human} that
DQNs are more successful with smaller action spaces.


Comparing Scenario 1 with 2, i.e., the dotted lines with the solid lines in Fig.~\ref{fig:sim},
we note that the performance of DQN1, which only uses scalar effective channel powers, is comparable with the baselines in Scenario 2, which use vectorized local effective channels for MRC.
Also, with higher $\rho$ values, the DQNs experience faster convergence, which is particularly noticeable in DQN1. 
The fluctuation of the DQN plots occurs due to the $\epsilon$-greedy policy, which explores random action selection occasionally to avoid getting trapped in local optima. Overall, in each case, we see that our MDP-based algorithms obtain significant performance improvements, emphasizing the benefit of our multi-agent DRL method.

%% file: conc.tex
\section{Conclusion}
\label{sec:conc}

We developed a novel methodology for uplink multi-IRS-assisted multi-cell systems. 
Due to temporal channel variations and difficulties of channel acquisition, we considered that BSs only acquire scalar effective channel powers from a subset of UEs.
We developed an information-sharing scheme among neighboring BSs and proposed a dynamic control scheme based on multi-agent DRL, in which 
each BS acts as an agent and adaptively designs
its local UE powers, local IRS beamformer, and its combiners.
Through numerical simulations, we verified that our algorithm outperforms conventional baselines.

%% file: main.bbl
\begin{thebibliography}{10}
\providecommand{\url}[1]{#1}
\csname url@samestyle\endcsname
\providecommand{\newblock}{\relax}
\providecommand{\bibinfo}[2]{#2}
\providecommand{\BIBentrySTDinterwordspacing}{\spaceskip=0pt\relax}
\providecommand{\BIBentryALTinterwordstretchfactor}{4}
\providecommand{\BIBentryALTinterwordspacing}{\spaceskip=\fontdimen2\font plus
\BIBentryALTinterwordstretchfactor\fontdimen3\font minus
  \fontdimen4\font\relax}
\providecommand{\BIBforeignlanguage}[2]{{%
\expandafter\ifx\csname l@#1\endcsname\relax
\typeout{** WARNING: IEEEtran.bst: No hyphenation pattern has been}%
\typeout{** loaded for the language `#1'. Using the pattern for}%
\typeout{** the default language instead.}%
\else
\language=\csname l@#1\endcsname
\fi
#2}}
\providecommand{\BIBdecl}{\relax}
\BIBdecl

\bibitem{zhang2020prospective}
J.~Zhang, E.~Bj{\"o}rnson, M.~Matthaiou, D.~W.~K. Ng, H.~Yang, and D.~J. Love,
  ``Prospective multiple antenna technologies for beyond {5G},'' \emph{IEEE J.
  Sel. Areas Commun.}, vol.~38, no.~8, pp. 1637--1660, 2020.

\bibitem{hosseinalipour2020federated}
S.~Hosseinalipour, C.~G. Brinton, V.~Aggarwal, H.~Dai, and M.~Chiang, ``From
  federated to fog learning: Distributed machine learning over heterogeneous
  wireless networks,'' \emph{IEEE Commun. Mag.}, 2020.

\bibitem{8811733}
Q.~{Wu} and R.~{Zhang}, ``Intelligent reflecting surface enhanced wireless
  network via joint active and passive beamforming,'' \emph{IEEE Trans.
  Wireless Commun.}, vol.~18, no.~11, pp. 5394--5409, 2019.

\bibitem{9013288}
H.~{Guo}, Y.~{Liang}, J.~{Chen}, and E.~G. {Larsson}, ``Weighted sum-rate
  maximization for intelligent reflecting surface enhanced wireless networks,''
  in \emph{Proc. IEEE Glob. Commun. Conf.}, 2019, pp. 1--6.

\bibitem{pan2020multicell}
C.~Pan, H.~Ren, K.~Wang, W.~Xu, M.~Elkashlan, A.~Nallanathan, and L.~Hanzo,
  ``Multicell {MIMO} communications relying on intelligent reflecting
  surfaces,'' \emph{IEEE Trans. Wireless Commun.}, 2020.

\bibitem{8990007}
H.~{Zhang}, B.~{Di}, L.~{Song}, and Z.~{Han}, ``Reconfigurable intelligent
  surfaces assisted communications with limited phase shifts: How many phase
  shifts are enough?'' \emph{IEEE Trans. Veh. Technol.}, vol.~69, no.~4, pp.
  4498--4502.

\bibitem{zhi2020uplink}
K.~Zhi, C.~Pan, H.~Ren, and K.~Wang, ``Uplink achievable rate of intelligent
  reflecting surface-aided millimeter-wave communications with low-resolution
  {ADC} and phase noise,'' \emph{arXiv:2008.00437}, 2020.

\bibitem{9162705}
L.~{Feng}, X.~{Que}, P.~{Yu}, W.~{Li}, and X.~{Qiu}, ``{IRS} assisted multiple
  user detection for uplink {URLLC} non-orthogonal multiple access,'' in
  \emph{IEEE Conf. Comput. Commun. Workshop}, 2020, pp. 1314--1315.

\bibitem{zhang2020intelligent}
S.~Zhang and R.~Zhang, ``Intelligent reflecting surface aided multi-user
  communication: Capacity region and deployment strategy,''
  \emph{arXiv:2009.02324}, 2020.

\bibitem{cui2014coding}
T.~J. Cui, M.~Q. Qi, X.~Wan, J.~Zhao, and Q.~Cheng, ``Coding metamaterials,
  digital metamaterials and programmable metamaterials,'' \emph{Light: Science
  \& Applications}, vol.~3, no.~10, p. e218, 2014.

\bibitem{wang2020channel}
Z.~Wang, L.~Liu, and S.~Cui, ``Channel estimation for intelligent reflecting
  surface assisted multiuser communications: Framework, algorithms, and
  analysis,'' \emph{IEEE Trans. Wireless Commun.}, 2020.

\bibitem{marinescu2017prediction}
A.~Marinescu, I.~Dusparic, and S.~Clarke, ``Prediction-based multi-agent
  reinforcement learning in inherently non-stationary environments,'' \emph{ACM
  Trans. Auton. Adapt. Syst.}, vol.~12, no.~2, pp. 1--23, 2017.

\bibitem{LTEstandard}
{3GPP TS 36.211}, ``{LTE}: Evolved universal terrestrial radio access (e-utra):
  Physical channels and modulation,'' vol. V14.2.0 Release 14, 2017.

\bibitem{mo2017channel}
J.~Mo, P.~Schniter, and R.~W. Heath, ``Channel estimation in broadband
  millimeter wave {MIMO} systems with few-bit {ADC}s,'' \emph{IEEE Trans.
  Signal Process.}, vol.~66, no.~5, pp. 1141--1154, 2017.

\bibitem{ge2020deep}
J.~Ge, Y.-C. Liang, J.~Joung, and S.~Sun, ``Deep reinforcement learning for
  distributed dynamic {MISO} downlink-beamforming coordination,'' \emph{IEEE
  Trans. Commun.}, vol.~68, no.~10, pp. 6070--6085, 2020.

\bibitem{mnih2015human}
V.~Mnih, K.~Kavukcuoglu \emph{et~al.}, ``Human-level control through deep
  reinforcement learning,'' \emph{nature}, vol. 518, no. 7540, pp. 529--533,
  2015.

\bibitem{au2007performance}
C.~K. Au-Yeung and D.~J. Love, ``On the performance of random vector
  quantization limited feedback beamforming in a {MISO} system,'' \emph{IEEE
  Trans. Wireless Commun.}, vol.~6, no.~2, pp. 458--462, 2007.

\bibitem{ieeep80216}
``{IEEE P802.16m}-2008 draft standard for local and metropolitan area
  network,'' \emph{IEEE Standard 802.16m}, 2008.

\bibitem{sklar2001digital}
B.~Sklar \emph{et~al.}, \emph{Digital communications: fundamentals and
  applications}, 2001.

\end{thebibliography}
